\documentclass[10pt,conference,colorlinks,linkcolor=blue,citecolor=blue,urlcolor=blue]{IEEEtran}

\IEEEoverridecommandlockouts
\usepackage{cite}
\usepackage{amsmath,amssymb,amsfonts}
\usepackage{graphicx}

\usepackage{caption}
\usepackage{subcaption}

\usepackage{textcomp}
\usepackage{xcolor}

\usepackage{soul} % added to handle comments
\usepackage{courier} % new font
\usepackage{tcolorbox} % box
\usepackage{xcolor} % box
\usepackage{hyperref} 

\usepackage{multirow}
\usepackage{array}
\usepackage{ragged2e}

\usepackage{algorithm}
\usepackage{algorithmicx}
\usepackage{algpseudocode}

\def\BibTeX{{\rm B\kern-.05em{\sc i\kern-.025em b}\kern-.08em
    T\kern-.1667em\lower.7ex\hbox{E}\kern-.125emX}}

\begin{document}

\title{Do AI assistants help students write formal specifications? A study with ChatGPT and the B-Method.\\


\thanks{The study received funding from the Department of Computer Science at the University of Luxembourg, under the AI4Education project.}
}
 
% 
% 
% \author{
%     \IEEEauthorblockN{Author One\IEEEauthorrefmark{1}, Author Two\IEEEauthorrefmark{2}}
%     \IEEEauthorblockA{
%         \IEEEauthorrefmark{1}Department of X, Institution Y, City, Country\\
%         Email: author1@example.com
%     }
%     \IEEEauthorblockA{
%         \IEEEauthorrefmark{2}Department of X, Institution Y, City, Country\\
%         Email: author2@example.com
%     }
%     \and
%     \IEEEauthorblockN{Author Three\IEEEauthorrefmark{3}, Author Four\IEEEauthorrefmark{4}}
%     \IEEEauthorblockA{
%         \IEEEauthorrefmark{3}Department of X, Institution Y, City, Country\\
%         Email: author3@example.com
%     }
%     \IEEEauthorblockA{
%         \IEEEauthorrefmark{4}Department of X, Institution Y, City, Country\\
%         Email: author4@example.com
%     }
% }
% 

\author{

\IEEEauthorblockN{Alfredo Capozucca\IEEEauthorrefmark{1}, Daniil Yampolskyi\IEEEauthorrefmark{1}, Alexander Goldberg\IEEEauthorrefmark{1}}

\IEEEauthorblockA{
	\IEEEauthorrefmark{1}
	\textit{Department of Computer Science} \\
	\textit{University of Luxembourg}\\
	Esch-sur-Alzette, Luxembourg \\
	alfredo.capozucca@uni.lu, \\ daniil.yampolskyi.002@student.uni.lu,\\ alexander.goldberg@ext.uni.lu
}
\and

\IEEEauthorblockN{Maximiliano Cristi\'a\IEEEauthorrefmark{2}}

%\IEEEauthorblockN{Maximiliano Cristiá}
\IEEEauthorblockA{
	\IEEEauthorrefmark{2}
	\textit{Universidad Nacional de Rosario}\\ and \textit{CIFASIS}\\
	Rosario, Argentina \\
	cristia@cifasis-conicet.gov.ar
}

}

% \author{
%     \IEEEauthorblockN{Author 1\IEEEauthorrefmark{1}, Author 2\IEEEauthorrefmark{2}, Author 3\IEEEauthorrefmark{3}, Author 4\IEEEauthorrefmark{4}}
%     \IEEEauthorblockA{\IEEEauthorrefmark{1}Department of Something, University A, Country}
%     \IEEEauthorblockA{\IEEEauthorrefmark{2}Department of Something, University B, Country}
%     \IEEEauthorblockA{\IEEEauthorrefmark{3}Department of Something, University C, Country}
%     \IEEEauthorblockA{\IEEEauthorrefmark{4}Department of Something, University D, Country}
% }

\maketitle

\begin{abstract}
This paper investigates the role of AI assistants, specifically OpenAI’s ChatGPT, in teaching formal methods (FM) to undergraduate students, using the B-method as a formal specification technique. While existing studies demonstrate the effectiveness of AI in coding tasks, no study reports on its impact on formal specifications. We examine whether ChatGPT provides an advantage when writing B-specifications and analyse student trust in its outputs. Our findings indicate that the AI does not help students to enhance the correctness of their specifications, with low trust correlating to better outcomes. Additionally, we identify a behavioural pattern with which to interact with ChatGPT which may influence the correctness of B-specifications.

\end{abstract}

\begin{IEEEkeywords}
requirements specification, formal methods, B-Method, AI assistants, ChatGPT 
\end{IEEEkeywords}

% % how to place comments

% Main contributors
% \fred{Test}
% \dani{Test3}

% Reviewers
% \maxi{Test2}
% \alex{Test4}

\section{Introduction}

% What is the context?
We are part of the community that believes it is important to teach formal methods (FM) to computer scientists~\cite{Broy_2024_FMedu} and that FM should be taught at the undergraduate level~\cite{Dongol_2024_FMedu}.   
One possibility is to teach FM for formalising requirements. This can be taught to undergraduate students who are already familiar with the fundamental concepts of software engineering. By the end of the course students should be able to write the formal specification for some given requirements.

% What is the concern?
It is a fact that students in higher-education use AI assistants~\cite{guardian2024aiessays,hepi2024policyai}. 
As educators, we are concerned with the level of performance that AI assistants may have in writing formal specifications, and in particular, how this may impact expected learning outcomes.  

% What is the motivation & % Why this is important
Evidence has shown that AI assistants can correctly complete coding tasks from introductory programming courses with minimal effort~\cite{Finnie-Ansley_Codex_2022_ASE,Deny_2023_AIA_introprog,Savelka_2023_AIA_introprog}. However, to the best of our knowledge, no study has investigated the degree to which AI assistants can provide help for novice students with writing formal specifications. Knowing if AI assistants are able to produce \emph{correct} formal specifications is highly relevant for educators when deciding how to design, deliver, and assess courses.

In this paper we examine if using OpenAI's ChatGPT~\cite{openai2024chatgpt} (referred to as the AIA) provides an advantage to undergraduate students when tasked with writing formal specifications using the B-method~\cite{schneider2001bmethod}. Moreover, we analyse what types of interactions lead to the maximisation of this advantage. In addition, we investigate whether students perceive an advantage when using the AIA to write their B-specifications. To do this, we designed a user study, inspired by the work of Perry et al.~\cite{Perry_2023}.

We designed the study to be coupled with a course that reunites all the required characteristics, but without causing any disruptions to its delivery and minimising any overhead for instructors. This not only ensures that participants represent the main targeted population, but also eases its reproducibility. We centred our study on three research questions:

% What we want to know
\begin{itemize}
%\item Do AI assistants help students to write \emph{``correct''} formal specifications?
\item \textbf{RQ1}: Does the AIA help students to write B-specifications?
%\item Do students trust AI assistants to write \emph{``correct''} formal specifications? 
\item \textbf{RQ2}: Do students trust the AIA to write B-specifications? 
\item \textbf{RQ3}: How do students’ behaviour when interacting with the AIA affect the degree of correctness of their B-specification?
\end{itemize}

% What we found
We found that using the AIA to write B-specifications does not help students perform better than when it is absent. Nevertheless, there is some, albeit weak, evidence that the AIA may help students specify B-abstract machine's state variables and find (but not write) the machine's operations (Section \ref{res-rq1}). 
We also found that most students had a low level of trust in the AIA. This lack of trust confirms that little advantage can be obtained from using the AIA. Students who expressed low level of trust are correlated with those who produced better B-specifications. (Section \ref{res-rq2}). The analysis of the participants' interactions with the AIA suggests the existence of a behavioural pattern that may lead to an enhancement of the correctness of the B-specification (Section \ref{res-rq3}). 

%These findings are based on two iterations of the study. 

\section{Background \& Related Works}

\subsection{The B-method}
The B-Method~\cite{Abrial_1996} is a formal method belonging to the category of state-based formalisms. The central idea of this method is to use the notion of an abstract machine to specify the possible states of a system. The transitions from one state to another are defined by the machine's operations. An abstract machine contains invariants, which describe properties of interest over the machine. A machine is said to be consistent when the invariant holds in every state. The B-language is used to specify the abstract machine, its invariants, and operations. The B-method also includes a process known as refinement, which aims to gradually transform an abstract machine into a concrete implementation, while ensuring that the invariants are preserved at each step of the transformation. The B-method is a success story for the FM community, not only for the research progress it has brought to the field but also for its proven track record in the industrial sector~\cite{Butler_2020_B_method_examples}.

\subsection{AIA and FM in education.}

We did not (at the time of writing this article) find any work reporting on how AIAs can help students to write formal specifications, or how it can be used to support learning of FM within computer science education. 

The closest work we found reporting on how AIAs can support FM is related to formal program verification~\cite{Janssen_2024_ChatGPT_loopinvariant}. This work analyses the capacity of ChatGPT (same AIA as ours) to generate valid and useful loop invariants for C programs. This work deals with aspects of FM related to formal verification of code (C programs, in this case), whereas our objective is to train students to formally specify requirements. Clearly, two different use cases of FM.

\section{Methods}

% FRED - GAINED SPACE
This section outlines the study design by detailing the composition of the participant pool,
the procedure to assess the correctness of B-specifications, the activities performed by participants, and the questions they were asked. Moreover, the procedure used when analysing the participants' prompts is described. Finally, this section also includes a comprehensive explanation of the experimental protocol, in order to ensure the reproducibility of the study, along with a presentation of the study's compliance with ethical regulations.

\subsection{Study design}

Our study is designed as a single group pretest-posttest quasi-experiment~\cite{trochim2015research}. In both tests we measure each participant's capacity to write formal specifications using the B-Method. This capacity is determined based on the correctness level (dependent variable) of the B-specification submitted by each participant as a result of having performed a guided activity. The criterion used to determine the correctness level of a B-specification is detailed in Section \ref{corr-cri}.

Before the pretest takes place, each participant receives training on formal requirements specification using the B-Method. This training is part of a course from which we recruit the participants for the study. Details about the course and participants are given in Section \ref{participants-pool}.

% Redundant information
% given in the final term of a 3-year undergraduate programme in computer science. This training consists of 23 clock hours, spread evenly over 7 weeks. The learning outcomes of the training aim at enabling participants to \emph{know} the purpose of formal specifications, and to \emph{apply} the B-Method to formalise requirements. Over these 7 weeks, participants are introduced to the role of formal methods in the software engineering field, refreshed on the notions of state machines, and instructed on how to specify \emph{consistent} state machines using the B language. Participants are requested to use the Atelier B tool\footnote{\href{https://www.atelierb.eu/en/atelier-b-support-maintenance/download-atelier-b/}{Atelier B 4.7.1, Community edition}.} to write their B-specifications.      

%The notion of consistency (i.e. the state machine should satisfy the invariant in its initial state, and any operation, once executed, preserves the invariant) is incrementally presented, explained and exercised multiple times when presented presented multiple  

The B-specification, used to measure the level of correctness during the pretest, is written by each participant using the Atelier B tool\footnote{\href{https://www.atelierb.eu/en/atelier-b-support-maintenance/download-atelier-b/}{Atelier B 4.7.1, Community edition}.} and without any assistance other than the class notes, examples, and reference documents provided during the training.

The treatment we provide to participants consists in utilising the AIA to write a subsequent B-specification. This AIA is provided as a new resource that participants can use on top of the previously available ones; the Atelier B tool, class notes, examples, and reference documents provided during the training. This second B-specification is the result of performing a second guided activity, but it is similar in terms of objectives and complexity to the first one. Guided activities are described in Section \ref{guided-act}.

The second B-specification is used to measure the level of correctness corresponding with the posttest. The variance of the correctness scores between the pretest and posttest is used to determine to what extent the AIA helps to write B-specifications, in line with \textbf{RQ1}. 

% Post activity survey: this survey is in line with RQ2
We ask participants to answer a questionnaire immediately following the submission of their second B-specification. This structured questionnaire (detailed in Section \ref{post-act-quest}) is designed to understand the participant's perception in terms of (p.1) how helpful the AIA is for specifying a B-specification, (p.2) their level of confidence that the submitted B-specification is correct, and (p.3) how much of this confidence is due to the help provided by the AIA during the process of writing the actual B-specification. 

The degree to which the participant perceived the AIA to be useful (i.e. part (p.1) of the questionnaire) is compared with the level of correctness of the submitted solution. This allows us to correlate how the believed help of the AIA relates with the correctness of the submitted B-specification. These findings are used to answer \textbf{RQ1}. 

Parts (p.2) and (p.3) of the questionnaire seek to ascertain the participant's level of confidence in the correctness of the provided B-specification, and how much of this confidence is due to the help of the AIA. The answers collected in these parts of the questionnaire are used to determine to what extent the participants trust the help provided by the AIA when writing a B-specification, thus contributing to answering \textbf{RQ2}.

% Logs of interaction with AIA: these logs are in line with RQ3.
Along with the B-specification, we ask each participant to submit their interactions with the AIA. These interactions contain both the prompt entered by the participant and the response generated by the AIA. Knowing how each participant has interacted with the AIA during the process of writing the B-specification allows us to recognise how much of an AIA-generated answer was \emph{reused} by a participant to produce the submitted B-specification. This metric provides quantitative data that is used to determine the level of trust of the participant in the AIA, in line with \textbf{RQ2}. 
 
It is worth clarifying how the reuse of an AIA-generated answer by a participant is determined. The \emph{reuse} of an AIA-generated answer by a participant may vary, from a verbatim copy-paste to a version modified with enhancements to its correctness. We use the notion of \emph{normalised edit distance} (NED)~\cite{Marzal_1993_NED} between the submitted B-specification and the closest AIA-generated answer, across all prompts, as a metric to determine how much of the AIA's responses a participant has reused in their submitted B-specification. This metric is used to corroborate the level of trust expressed by participants via the questionnaire. 

Consistent data should show that participants who trust in the AIA\footnote{The participant is confident that the correctness of the submitted B-specification is due to the help of the AIA.} should have a small NED. The triangulation between the qualitative data collected via the questionnaire and the quantitative data obtained by measuring NED helps to provide a better understanding of how much the participants trust the AIA, thus helping to answering \textbf{RQ2}.

We also use the prompt interactions provided by each participant to determine the \emph{type of prompts} used by participants, and how (i.e. when and number of times) each of them are used over the process of writing the B-specification. The analysis of the prompt interactions (detailed in Section \ref{prompt-analysis-proc}) aims to identify insights on the types of prompts that may help producing correct B-specifications. This analysis should help answering \textbf{RQ3}.

\subsection{Recruitment and participant pool}\label{participants-pool}

% The course
We recruited undergraduate students who signed up for an optional course on formal specification and automated verification. This course is part of a 3-year undergraduate programme in computer science. The course is placed in the last term of the last year of the programme. At this point students have already acquired the fundamentals of computer science (e.g. set theory, propositional and predicate logic, sate machines), software engineering, and proven programming experience in several paradigms (imperative, objected oriented, and functional programming). The course lasts 14 weeks, with a weekly load of 3 wall-clock hours. The first 7 weeks of the course are spent focusing on the B-method, whereas the last 7 weeks concentrate on automated verification (using a different state-based formalism and tool support). For the purpose of the study, we are only concerned with the part covering the B-method, only. The learning outcomes of this part aim at enabling participants to \emph{know} the purpose of formal specifications, and to \emph{apply} the B-Method to formalise requirements. Over these 7 weeks, participants are introduced to the role of formal methods in the software engineering field, refreshed on the notions of state machines, and instructed on how to specify \emph{consistent} state machines using the B language. Participants use the Atelier B tool to write their B-specifications.

% Assumptions about the participants
Recruiting participants from this course ensured (based on the programme curricula) that they did not have any previous knowledge nor experience in formal specification using the B-method. Thus, we can say with high confidence that the acquired knowledge and experience on the B-method at the moment of participating in the experiment was gained during the first 7-weeks period of the course.  

% Campaigns
We organised two recruitment campaigns, corresponding with the 2023 and 2024 editions of the course (referred to as iteration 1 and 2, respectively). Course-wise, nothing changed between each iteration. Each recruitment campaign was led by the instructors of the course and officially launched the day that feedback of the first guided activity (which is worth 30\% of the course's final mark) was given to students. On this day a 30 minute information session was given and printed versions of the experiment notice and authorisation form were distributed. Both documents were approved by our institution's Ethical Review Panel (ERP). We did not offer students any incentive (of any kind) to participate in the study. Ultimately, we recruited 6 (out of 7) students for the first iteration, and 10 (out of 19) students for the second. Table \ref{tab:1} contains a summary of the demographics of our participants.

\begin{table}
\caption{Summary of participant demographics per iteration.}
\label{tab:1}
\centering
\setlength\extrarowheight{3pt}

\begin{tabular}{|l|c|c|c|}

\hline
\textbf{Demographics} & \textbf{Cohort} & \textbf{Iteration 1} & \textbf{Iteration 2} \\ 
\hline
% \multirow[t]{2}{2.5cm}{Gender} & Male & 83\% & 70\% \\ 
%  & Female & 17\% & 30\% \\ 
% \hline
% \multirow[t]{2}{2.5cm}{Age} & 18-24 & 67\% & 90\% \\ 
%  & 25-34 & 33\% & 10\% \\ 
% \hline
% \multirow[t]{2}{2.5cm}{Country of birth} & EU & 83\% & 70\% \\ 
%  & Non-EU & 17\% & 30\% \\ 
% \hline
% \multirow[t]{2}{2.5cm}{Mother tongue} & EN & 17\% & 10\% \\ 
%  & Other & 83\% & 90\% \\ 
% \hline
% \multirow[t]{4}{2.5cm}{Programming \newline experience \newline outside of studies \newline (in years)} 
%  & 0-1 & 0\% & 60\% \\ 
%  & 1-2 & 33\% & 20\% \\ 
%  & 2-5 & 67\% & 10\% \\ 
%  & 5+ & 0\% & 10\% \\ 
% \hline
\multirow[t]{4}{2.5cm}{ChatGPT Usage} 
 & Very often & 17\% & 10\% \\ 
 & Often & 33\% & 30\% \\ 
 & Regularly & 17\% & 30\% \\ 
 & Few times & 33\% & 30\% \\ 
 & Never & 0\% & 0\% \\ 

\hline
\multirow[t]{5}{2.5cm}{Activities with ChatGPT}
 & Info retrieval & 21\% & 19\% \\
 & Programming & 29\% & 26\% \\  
 & Writing essays & 36\% & 22\% \\ 
 & Other & 14\% & 33\% \\
\hline
\multirow[t]{5}{2.5cm}{Premium version}
 & Yes & 17\% & 30\% \\
 & No & 83\% & 70\% \\  
\hline
\end{tabular}

\end{table}

What stands out immediately upon viewing the demographics is that all participants had already used the AIA (i.e. ChatGPT\footnote{Recall, OpenAI released ChatGPT in November 2022, and the first iteration of this experiment took place in May 2023.}), mainly for programming and writing activities. Several participants even acknowledged using the premium version of the service, which may be interpreted as a confidence in the quality offered by the service (hence their willingness to invest in it). 

%\fred{Indicate examples of Other activities done with ChatGPT.}

Participants also indicated having used the AIA for writing/correcting emails, and generating random data.

%\fred{There are not actual results that discriminate by Mother tongue or Country of birth, so such as information is pointless.}
%inexperienced programmers to those having a stack of mastered languages, all were from different countries. 

% FRED - CHECK IF THIS INFO IS NOT REPEATED IN SECTION "THREATS TO VALIDITY"
The number of participants in each iteration is low. Thus, the results obtained so far are better suited for exploratory analysis rather than definitive conclusions. Under the current conditions\footnote{Course organisation (i.e. optional, and with two B-related guided activities) and participants' profiles (undergrad students with zero knowledge of the B-method).} it is quite unlikely that we could recruit a larger number of participants (i.e. at least 30) in future iterations. Nevertheless, replicating the study over time allows us to determine whether found insights and trends remain valid or have changed.

\subsection{Correctness assessment procedure}\label{corr-cri}

The correctness of each submitted B-specification was assessed by the 2 instructors in charge of delivering the course where the experiment took place. Each instructor individually provided a correctness percentage ranging from 0\% to 100\%, with possible values rounded to the nearest 10\% for each dimension making up the grading criteria. 

These dimensions are aligned with the parts that compose a B-method's \emph{abstract machine} (actual representation of a B-specification). A correct B-specification is expected to fulfil each of these parts with information that formally captures the given informal requirements. These dimensions along with their weight distribution are:

\begin{itemize}

\item \textit{State variables (10\%):} the extent to which the abstract machine's state variables capture the concepts of the domain of discourse, along with their structure, data values, and relationships.

%For example, a system aimed at handling banking accounts, should have concepts like \emph{AccountNumber} and \emph{Balance} related to each other to indicate that a bank account has a balnace. Moreover, as the bank handles multiple accounts, the system should have a state variable like \emph{Accounts} that allows to collect many bank accounts. The structure of each concept and variable depends on the provided requirements. 

\item \textit{Initialisation (5\%):} the extent to which the abstract machine's state variables are initialised.

\item \textit{Abstraction level (15\%):} the extent to which the defined abstract data types match the level of detail required to specify the given requirements. 

\item \textit{Operation specification (40\%):} the extent to which the abstract machine provides not only the operations that satisfy the required functional requirements, but also the level of detail (i.e. main and alternative cases are specified) of each operation.

\item \textit{Invariants (15\%):} the extent to which the properties of the abstract machine are discovered and specified. Notice that proof of invariant discharge is not part of the assessment covered by this dimension. 

\item \textit{Type checking (15\%):} the extent to which the abstract machine adheres to the B-language's syntax and uses the language's expressiveness power.

\end{itemize}

The weight assigned to each dimension is proportional to the time spent during the course explaining and working on the topic. After the individual assessment, the 2 instructors met to check their marking and to reach an agreement in case of discrepancies. We chose this strategy to increase fairness and consistency among the participants' marks.

% The assessment of this dimension is done automatically using the Atelier B tool as it provides a feature to perform syntax analysis and type checking of a B specification\footnote{Also referred to as \emph{abstract machine} in the B-method.}.

\subsection{Guided activities}\label{guided-act}
Students who decide to participate in the study perform two guided activities. In both activities, participants are given a plain English statement describing the requirements of software to be developed. Participants then have to write a B-specification fulfilling these requirements using the Atelier-B software tool. We give participants up to 120 minutes to complete the activity. During this time participants have access to the class notes, examples, and reference documents provided during the training. The complexity of the requirements were not only the same between both activities, but also very similar to the exercises done during the training. 

% OLD
% Participants have to export the specification using Atelier-B's features\footnote{Only specifications that can be successfully imported in Atelier-B are considered for assessment.} and submit it via the official learning management system (LMS) of our institution\footnote{This allows us to comply with the institution's data regulations.} The environmental conditions were the same between both activities: i.e. same room, hardware, software, complexity of the requirements, and time to complete the activity.

Participants have to submit the specification via the official learning management system (LMS) of our institution\footnote{This allows us to comply with the institution's data regulations.} The environmental conditions were the same between both activities: i.e. same room, hardware, software, complexity of the requirements, and time to complete the activity. 

The only difference between the first and second guided activity, is that in the second participants are allowed to use the AIA, at their discretion, to write the B-specification\footnote{Using the AIA in the first guided activity was not allowed. The small number of participants made it possible to enforce this rule.}. We ask participants to try to reuse as much of the AIA-generated answers as possible, but to prioritise correctness above all else. Participants are requested to submit the complete logs of their interactions with the AIA along with their B-specification.

% Specifications were evaluated with the use of assessment criteria, that represent specification's properties, that involved the previously mentioned specification areas:

% \begin{itemize}

% \item	 \textit{Import} to prove the participant has some previous experience with Atelier B and correct export.

% \item	 \textit{Type checking} is expected that the specification is well typed.

% \item	 \textit{State variables} if the given variables are enough to capture the state of the system. Assesses participant’s capacity to digest the provided information.

% \item	 \textit{Invariants} if every state has a type.

% \item	 \textit{Initialisation} if every state variable correctly initialised.

% \item	 \textit{Abstraction Level} if the participant finds the right level of abstraction based on the given prompt describing the activity and types that are defined, as well as constants and/or parameters.

% \item	 \textit{Operation specification} is determined based on whether it covers non-only happy path, returns a meaningful message, and level of coverage of unhappy paths.

% \end{itemize}

\subsection{Post-activity questionnaire}\label{post-act-quest}

The post-activity questionnaire is designed to collect qualitative information from the participants related to the level of help (\textbf{RQ1}) provided by the AIA (questions q1-q8) during the process of writing the B-specification. We want to gain a finer view into the help provided by the AIA when writing specifications. We ask participants to indicate the level of help to \emph{identify what to specify} and to \emph{write} the actual specification. The questions are strongly aligned with the same dimensions used during the correctness assessment. We ask participants to indicate the level of help to find out and write \emph{operations, invariants,} and \emph{state variables}. For the dimension \emph{type checking} (adherence to the B-language) we only ask for the level of help to write the concerned parts of the specification previously indicated. 

The level of help for questions q1-q8 is indicated using a 5-point Likert scale that consists of: (0) Null, (1) Poor, (2) Average, (3) Good, and (4) Very good. Question q9 gives participants the possibility to indicate activities (and their level of help) that do not fall into the indicated dimensions. Question q10 asks participants to indicate their level of confidence about the correctness of the submitted B-specification. This confidence is indicated using a 5-point Likert scale that consists of:  (-2) Very low, (-1) Low, (0) Average (1) High, and (2) Very high.

The follow-up question, q11, is aimed at determining how much of the indicated confidence in q10 is due to the help of the AIA. A 5-point Likert is also provided to participants to let them answer this question. It consists of: (1) Very little, (2) Little, (3) Some, (4) Much, and (5) Very much. Participants' answers to questions q10 and a11 are used to understand how much trust in the AIA participants have, which contributes to answering \textbf{RQ2}.

The complete list of questions is provided in the replication package~\cite{capozucca_2024_13914857}.

\subsection{Prompt analysis}\label{prompt-analysis-proc}
We adapt Perry et al.~\cite{Perry_2023}'s taxonomy to categorise the prompts provided by participants. This taxonomy is described as follows:

\begin{itemize}
\item \textit{SPECIFICATION:} user provides a natural language description of requirements to be specified (e.g. \texttt{``menu is a collection of dishes that are available in the food truck (ex.: cheese burrito, chicken burrito, etc)''}).

\item \textit{INSTRUCTION:} user indicates to the AIA what B-language instruction to use (e.g. \texttt{``for the recordSale operation we can use SELECT instead of the normal IF THEN.''}).

\item \textit{QUESTION:} user asks the AIA a question (e.g. \texttt{``but what do you mean by the dot ?''}).

\item \textit{LANGUAGE:} user indicates the target language to use (e.g. \texttt{``write me a b specification for ...''}).

\item \textit{TEXT CLOSE:} normalised edit distance between prompt and \emph{question text}\footnote{It refers to a question that appears in the guided activity sheet which the participant has to answer.} is less than 0.25 (i.e. the strings are \textbf{very similar}). 

\item \textit{MODEL CLOSE:} normalised edit distance between prompt and the previous AIA's output is less than 0.4 (i.e. the strings are \textbf{similar}).

\item \textit{HELPER:} prompt includes examples of B-specifications.

\item \textit{PROBLEM:} user reports a problem with some parts of a B-specification (e.g. \texttt{``I get an error in Atelier b saying ...''}).

\item \textit{WARMUP:} prompt includes contextual information.

\item \textit{ROLE:} user instructs the AIA to play a particular role.

\item \textit{COMMAND:} user provides a command that is informative for the AIA (e.g. \texttt{``No, it should not be 1, it should be the same as before.''}). 

\end{itemize}

The process of categorising the prompts is done manually. It was executed by 3 independent researchers, who met at the end of the process to reach a consensus on the final categorisation. Notice that a same prompt may belong to multiple categories.

Having categorised prompts, allows us to calculate the average position of a type of prompt (i.e. category) over the dialogue of a particular participant. A position ranges from 0 to 1, where 0 is the beginning of the dialogue and 1 is the end of the dialogue. Thus, a prompt type with an average position falling in the range [0,0.33] means that prompts of this category are likely to be used at the beginning of the dialogue. Conversely, a type of prompt with average position in [0.66, 1] indicates that prompts in this category are used more towards the end. 

Knowing the average position of each type of prompt for each participant allows us to determine the overall position distribution of each type of prompt over participants. The distribution of the categories over the dialogue period allows us to find out if there exists any behavioural patterns in how these categories are used over the duration of a dialogue.

We assume that every AIA-generated answer produces an effect on the submitted B-specification. We reduce the impact of this assumption by only considering participants who have reused AIA-generated answers (i.e. NED $<$ 0.4) in their submitted B-specification. As reuse is a proxy metric for trust, participants with high reuse are considered as participants who believe they are being positively assisted by the AIA.

As the goal is to find trends that help to write ``good'' B-specifications, we further reduce the space of prompts to be analysed. We consider prompts from participants whose submitted B-specifications have a correctness level $\ge$ 50\%. Inspecting the prompts of this sub-set of participants should allow us to discover patterns that lead to producing correct B-specifications, if any.

% We use these sub-set of participants to find out trends among type of prompts. Dividing participants between those who have produced good enough B-specifications (i.e. correctness level $\ge$ 50\%) and those who have may allows us to discover patters that help producing correct specifications, as well as patterns that are drawbacks towards this end.

We perform a similar analysis to investigate the impact a type of prompt may have on the correctness of a B-specification. Using the prompt categorisation and correctness level of each participant, we calculate the contribution of each category to the obtained correctness score. Once we know the contribution of each type of prompt for each participant, we can determine how these contributions are distributed among this subset of participants. The average distribution of each type of prompt is used as an indicator to determine its impact on correctness.

\subsection{Protocol}\label{method-protocol}

To ensure the reproducibility and clarity of our study, we provide a step-by-step description of the protocol used to run the experiment.

\textbf{(1) Pre-settings:} Our participants are a sub-set of the students following the undergraduate course where the study is embedded. By the time the experiment is performed, each student (regardless of whether they want to participate in the study or not) has already completed the \textbf{first guided activity} as it is part of the course. The B-specifications submitted by students are assessed for correctness. The level of correctness of this specification represents 30\% of the student's final mark in the course. This provides sufficient motivation to ensure that every student who intends to pass the course participates in this guided activity. However, it is worth noting that for the sake of the study, we retain only the correctness assessment of students who agree to participate in the study.

\textbf{(2) Info session:} We present in 30 minutes what the study is about, how it is done, and how personal data will be processed. We clearly explain to the students that to participate in the study they must have a registered account to use the AIA. During the presentation we distribute the information notice and authorisation form that every student has to return duly signed to be able to participate in the study. We give students the time until the experiment starts (2 weeks later) to decide whether to participate or not.

% Remarks: 
% [Only for the experiment made in 2023]: participation in the experiment may allow a student to get 2 extra points in the final grade of the course. These 2 extra points are given based on the level of completeness  of each task given to the participant (write specification and answer questionnaires). 

% we expect students to sign up for the free version, but a question is asked to check whether they use the premium version or not.
% it is mentioned before the study is executed that "is a problem to depend on a flaky third party service”. This is a good mental preparation for students who will struggle to get access to the tool during the execution of the experiment.

\textbf{(3) Experiment kick off:} We recall the objectives of the experiment while collecting the duly signed agreement forms. We present the questions to be answered in the questionnaire following the activity. We ask participants to fill out a survey (15 minutes) meant to collect demographic information related to the goals of the study.

\textbf{(4) Second guided activity:} We distribute the activity sheet and ask students to proceed with its execution. This time students are allowed to use the AIA when doing the activity. This is the only difference with respect to the first guided activity they have already completed. 

\textbf{(5) Post-activity questionnaire:} Once a participant has completed the activity they are allowed to answer the follow-up questionnaire. We inform participants that to complete the questionnaire they may need between 15 and 20 minutes.

\subsection{Ethics}

%\fred{Include in this section the approval of the ER board to perform the study.}

We received the approval of our institution's ERP to perform the study. Students signed consent forms prior to the experiments, ensuring they were aware of how their personal data would be processed and used. Students were also informed that not participating in the study would not represent a disadvantage. We made explicit in the study's information notice that the learning outcomes to be acquired during the course are still guaranteed regardless of their participation in the study.

% To capture whether AI assistant helped to write “correct” formal specifications, participants first wrote assignments relying on the knowledge they had gained during the course. After that they were asked to use ChatGPT to write the specification. 

% Before the study was executed it was mentioned that it "is a problem to depend on a flaky third-party service”. Such mental preparation was for students who would struggle to get access to the tool during the execution of the experiment.

% Delving into the use of ChatGPT, participants were familiarised with the rules of use and basic guidelines. \dani{Note from education.experiment-chatGPT:} Participants were expected to sign up for the free version, but a question was asked to check whether they used the premium version or not. Participants were asked to log in a .txt document with the interaction they had with the assistant when writing the specification. 

% After the experiment activity, participants were asked to answer the follow up questionnaire \dani{Note from education.experiment-chatGPT:} which was presented at the very beginning of the session and contained questions regarding their evaluation of chatbot’s help.

% While students were motivated by the offer of 2 extra course points (for the first iteration), they were assured that their participation would not significantly impact their education. 

\section{Results}\label{results}

%\section{Formal specification analysis}

\subsection{RQ1: Does the AIA help students to write B-specifications?}\label{res-rq1}

Table \ref{tab:sum-corr} presents a summary of the level of correctness of the submissions. It reveals a decay in performance between the pretest and posttest, which appears in both iterations. This is a first hint towards the idea that the AIA does not help the students perform better.

\begin{table}
\caption{Summary of correctness for each guided activity.}
\label{tab:sum-corr}
\centering
\setlength\extrarowheight{3pt}

\begin{tabular}{|l|c|c|c|}

\hline
\multirow[c]{2}{2.5cm}{\textbf{Guided Activity}} & \multicolumn{3}{c|}{\textbf{Correctness}} \\
\cline{2-4}
 & \textbf{Average} & \textbf{Min} & \textbf{Max} \\
%& \textbf{Iteration 1} & \textbf{Iteration 2} \\ 
\hline
\hline
\multicolumn{4}{|c|}{\textbf{Iteration 1}} \\
%\hline
% & \textbf{Average} & \textbf{Min} & \textbf{Max} \\
\hline
First (pretest)  & 77,92\% & 45,00\% & 100,00\% \\
\hline
Second (posttest)  & 59,78\% & 34,78\% & 93,48\% \\
\hline
\hline
\multicolumn{4}{|c|}{\textbf{Iteration 2}} \\
\hline
% & \textbf{Average} & \textbf{Min} & \textbf{Max} \\
%\hline
First  (pretest)  & 73,33\% & 40,00\% & 100,00\% \\
\hline
Second (posttest)  & 70,88\% & 35,29\% & 100,00\% \\
\hline
\end{tabular}
\end{table}

% Result's analysis using Fig 1

\begin{figure}[htp]
    \centering
    % First subfigure
    \begin{subfigure}[b]{1.0\linewidth} 
        \centering
        \includegraphics[width=1.0\textwidth]{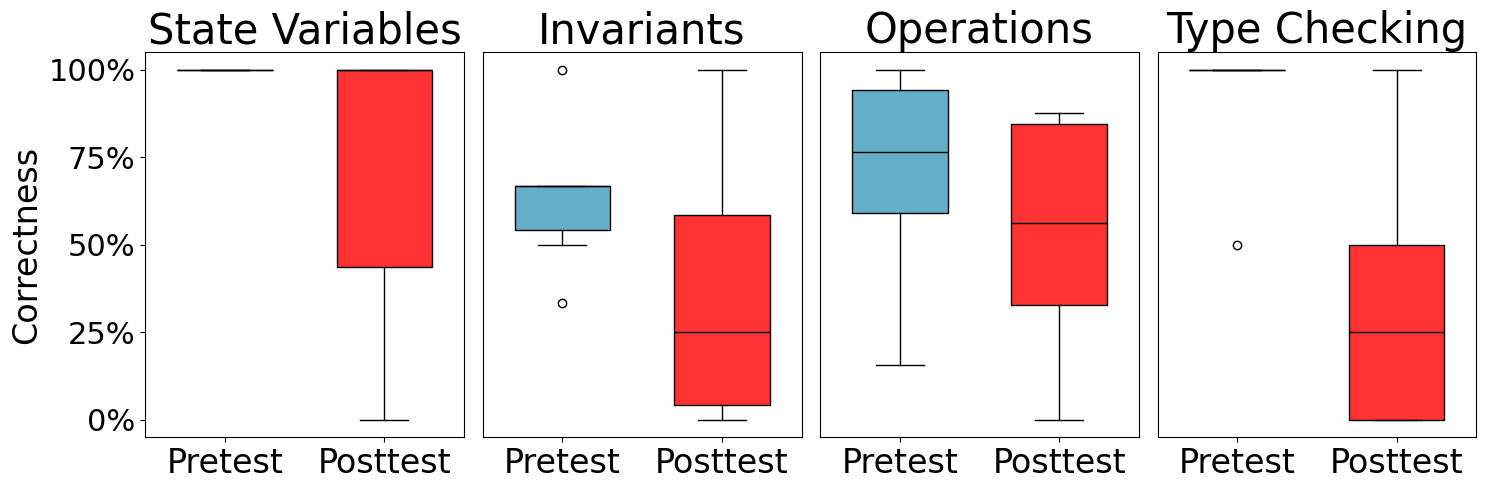} % Replace with your figure
         \caption{Iteration 1}
        \label{RQ1-prepost-tests-1}
    \end{subfigure}
    
    \vspace{0.2cm} % Adjust space between the figures
    
    % Second subfigure
    \begin{subfigure}[b]{1.0\linewidth} 
        \centering
        \includegraphics[width=1.0\textwidth]{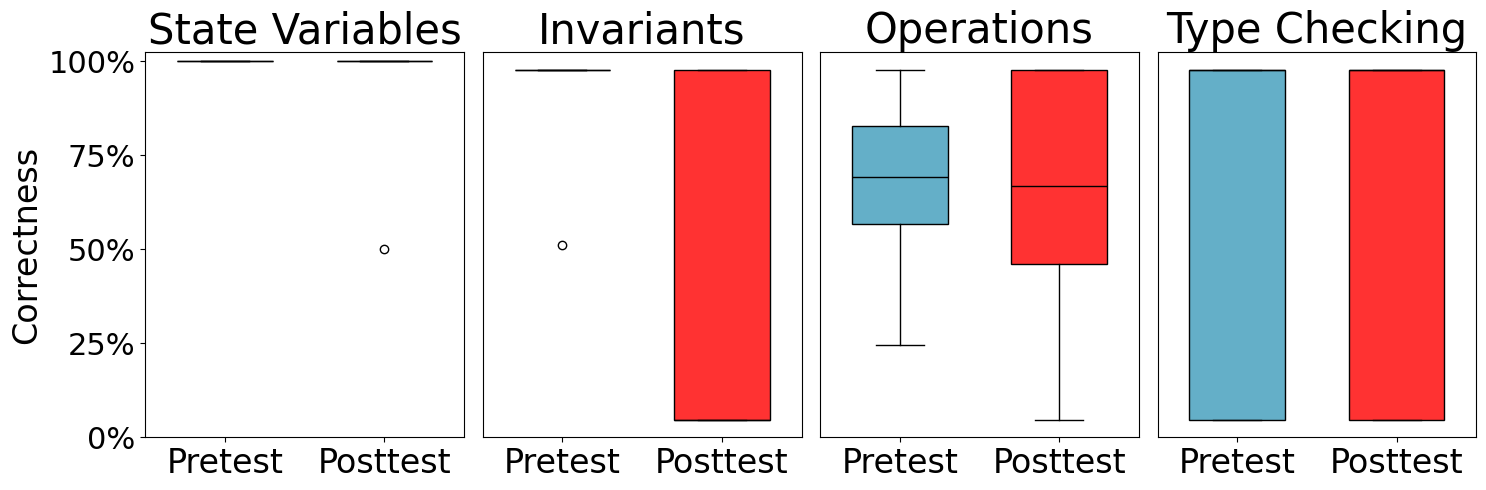} % Replace with your figure
        \caption{Iteration 2}
        \label{RQ1-prepost-tests-2}
    \end{subfigure}
     \caption{Correctness distribution of participants for each assessment criterion's dimension.}
     %\caption{Correctness distribution of participants for each assessment criterion's dimension. Results show that the use of the AIA did not help participants to increase the correctness of their B-specification on any dimension.}
     \label{RQ1-prepost-tests}
\end{figure}

Next, we zoom in from the overall performance of participants to investigate whether the AIA made a difference in any of the correctness assessment criterion dimensions. Figure \ref{RQ1-prepost-tests} shows how participants performed in each dimension\footnote{We remove the dimension ``initialisation'' from the figure as every participant achieved 100\% in the pretest and posttest in both iterations. Thus the plot-charts for this dimension do not add any meaningful information.} during the pretest and posttest. The results show that no increase in performance was observed in any dimension, in terms of median value, when using the AIA. However, it is worth noticing that iteration 2's results show not only an overall close median performance between the pretest and posttest, but also a better performance of some participants when the AIA was available. This evidence shows that the AIA may have helped when specifying the B-abstract machine's operations.     

When comparing the distributions between both iterations, results show an increase of the performance (median value) when using the AIA. These results may be due to advancements in the quality of the AIA\footnote{During the first iteration (May 2023), the unpaid version of the AIA was GPT-3.5, whereas in the second iteration (May 2024) it was GPT-4.}, improvements in the participants' abilities to utilise the AIA, or a combination of both.

% \begin{figure*}[t!]
%     \centering
%     \begin{subfigure}[t]{0.5\textwidth}
%         \centering
%         %\includegraphics[height=1.2in]{a}
%         \includegraphics[width=1\textwidth]{graphs/radarR1.png}
%         \caption{Iteration 1}
%     \end{subfigure}%
%     ~ 
%     \begin{subfigure}[t]{0.5\textwidth}
%         \centering
%         %\includegraphics[height=1.2in]{b}
%         \includegraphics[width=1\textwidth]{graphs/radarR2.png}
%         \caption{Iteration 2}
%     \end{subfigure}
%     \caption{Participants' average success rates in solving five shared properties of the specifications for each task (tasks with and without access to chatGPT) for the first and second iterations. Participants who used chatGPT to solve the specifications performed worse in both iterations}
%     \label{RQ1-Radar}
% \end{figure*}

\begin{figure*}[t!]
    \centering
    \begin{subfigure}[t]{1.0\linewidth}
        \centering
        \includegraphics[width=1\textwidth]{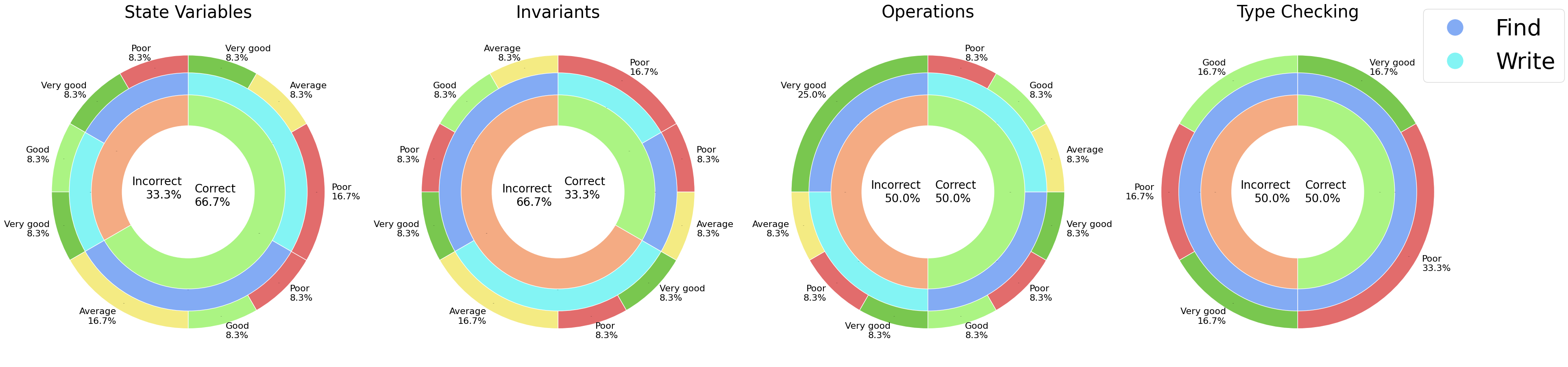}
        \caption{Iteration 1}
    \end{subfigure}
    
    \begin{subfigure}[t]{1.0\linewidth}
        \centering
        \includegraphics[width=1\textwidth]{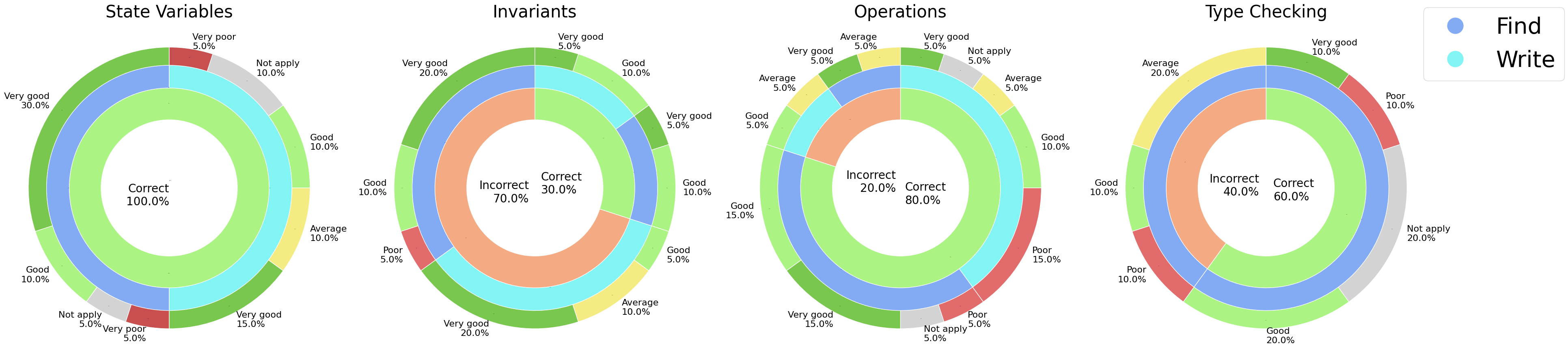}
        \caption{Iteration 2}
    \end{subfigure}
    \caption{Distribution of participant's beliefs regarding the help offered by the AIA to achieve a B-specification and their actual achieved performance. }
   \label{RQ1-corr-survey-boxCharts}
\end{figure*}

Producing a B-specification involves performing two activities: (i) to identify what to specify, and (ii) to formulate the actual specification. The help provided by the AIA to complete these activities, as perceived by participants (obtained via the post-activity questionnaire), contrasts with their achieved performance. Figure \ref{RQ1-corr-survey-boxCharts} shows the distribution of participants' answers concerning the quality of the help provided by the AIA, for each correctness dimension. The participants' answers are divided between those who provided a ``good'' B-specification\footnote{The correctness level required to attain this division is 50\%, which corresponds with the minimal passing mark of the course} (positive group), and those who did not (negative group). This discrimination allows us to have better visibility on how participants' beliefs align with their achieved performance. Beliefs held by the positive group in both iterations  are significant indicators of how helpful the AIA is for each task.

Each multi-level doughnut chart is read from the inside to the outside. The inner most ring of each chart corresponds to the group category. This category is divided into two segments: the positive group and the negative group (i.e. participants who provided a correct or an incorrect solution, respectively). The second ring corresponds to the AIA help category, which is divided into the segments \emph{find} and \emph{write}. The outermost ring is used to show the distribution of the participants answers for each segment of the middle ring.

% For example, the multi-level doughnut that corresponds to \emph{State variable} in the first iteration, shows that 66.7\% of the participants have correctly specified state variables. Among them, 8.3\% indicated that the level of help offered by the AIA to \emph{find} what to specify is poor. However, 16.7\% of the participants in the same group indicated that the level is poor regarding the help offered by the AIA to \emph{write} the actual specification.

% OLD
% \begin{figure*}[t!]
%     \centering
%     \begin{subfigure}[t]{0.5\textwidth}
%         \centering
%         %\includegraphics[height=1.2in]{a}
%         \includegraphics[width=1\textwidth]{graphs/trust_corr_distR1.png}
%         \caption{Iteration 1}
%     \end{subfigure}%
%     ~ 
%     \begin{subfigure}[t]{0.5\textwidth}
%         \centering
%         %\includegraphics[height=1.2in]{b}
%         \includegraphics[width=1\textwidth]{graphs/trust_corr_distR2.png}
%         \caption{Iteration 2}
%     \end{subfigure}
%     \caption{Distribution of participant's beliefs regarding the help offered by the AIA to achieve a B-specification and their actual achieved performance. }
%    \label{RQ1-corr-survey-boxCharts}
% \end{figure*}

For example, the multi-level doughnut that corresponds to \emph{state variable} in the first iteration, shows that participants in the positive group (66.7\%) perceived some (16.7\% average, and 8.3\% good) help from the AIA to identify such variables. However, their perception about the help of the AIA to write these variables was negative (16.7\% poor). The fact that participants in the negative group (33.3\%) perceived to have obtained good help (16.6\% combined) from the AIA to write state variables may confirm that actually the AIA is not helpful with writing state variables. However, the AIA may provide some help to identify them. Iteration 2's results allow us to confirm the trend that \textbf{the AIA may help with finding state variables}. However, the initial hint that the AIA does not help to write state variables is now weakening. Thus, based on the facts that (i) there were more participants in the second iteration, and (ii) the positive group was larger, it seems \textbf{the AIA may also be helpful for writing state variables}.

Iteration 1's results about \emph{invariants} show that participants in the positive group (33.3\%) perceived poor help from the AIA, both in finding and writing invariants. This trend changed in the second iteration, as participants in the positive group (30\%) found the AIA to be helpul with finding (15\%) and writing (15\%) invariants. While iteration 2's results may hint towards the AIA being helpful for specifying invariants, the facts that (i) the negative group was larger (70\%), and (ii) they also perceived help from the AIA, leads us to invalidate this impression. Thus, based on the results from both iterations, \textbf{the AIA does not appear to help with finding nor writing invariants}.

Results show an alignment between the participants' perceptions in both iterations regarding help from the AIA in relation with the specification of \emph{operations}. Most participants from the positive group (50\% and 70\% in iteration 1 and 2, respectively), found that the AIA was helpful for finding operations, but not for writing them. The fact that this trend is confirmed in both iterations for the positive group, and that the proportion of this group has increased in the second iteration, leads us to conclude that \textbf{the AIA seems to help with finding operations, but not to write them}.

For the \emph{type checking} dimension, participants are only asked to indicate the level of help offered by the AIA to \emph{find} what to specify\footnote{Recall that this dimension aims at assessing how much the provided B-specification adheres to the B-language's syntax and uses its expressive power. Thus, the need of the participant is to \emph{find} whether certain ``B-code'' is correct, regardless of whether it is AIA-generated or not.}. 

Iteration 1's results show that most participants from the positive group (50\%) found the ability of the AIA to help to be poor when finding (i) how to write well-formed B-expressions and (ii) which B-operator was the most suitable in such expressions. This belief is confirmed by the fact that participants from the negative group found the help offered by the AIA to be good in this regard.    

Iteration 2's results show a similar trend concerning participants' beliefs, but with some indicators that the AIA may provide some help in this dimension. These indicators are (i) the increase of participants in the positive group (10\% more than in iteration 1) and (ii) half of the participants from this group having found some help from the AIA. We prefer to be cautious and conservative, thus we prefer to state that \textbf{the AIA does not seem to help to write well-formed B-expressions} until more stronger evidence showing the contrary is found.

In summary, results from both iterations show that the use of the AIA \textbf{did not help participants} to increase their overall performance when writing B-specifications. As of today, writing a correct B-specification still depends more on the knowledge and skills of the student than the help obtained from the AIA. 

The evidence obtained from the study shows that using the AIA does not guarantee that a correct B-specification will eventually be produced. However, the same evidence shows that the AIA may provide some help during the process of producing the final B-specification, in particular finding what has to be specified.

\subsection{RQ2: Do students trust the AIA to write B-specifications?}\label{res-rq2}
%\section{Trust analysis}

While the first research question aims to understand how much help can be obtained from the AIA to write a B-specification, RQ2 aims to determine how much of the obtained help makes it into the final B-specification. Questions \emph{q10} and \emph{q11} of the post-activity questionnaire (see Section \ref{post-act-quest}) aim to collect information to help answer RQ2.

\begin{figure}[t!]%
    \centering
    \subfloat[\centering Iteration 1]{{\includegraphics[width=3.9cm]{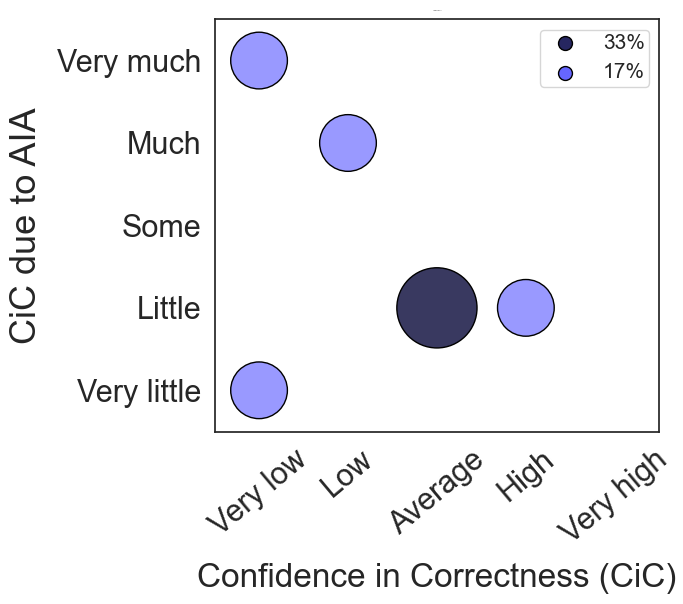} }}%
    \qquad
    \subfloat[\centering Iteration 2]{{\includegraphics[width=3.9cm]{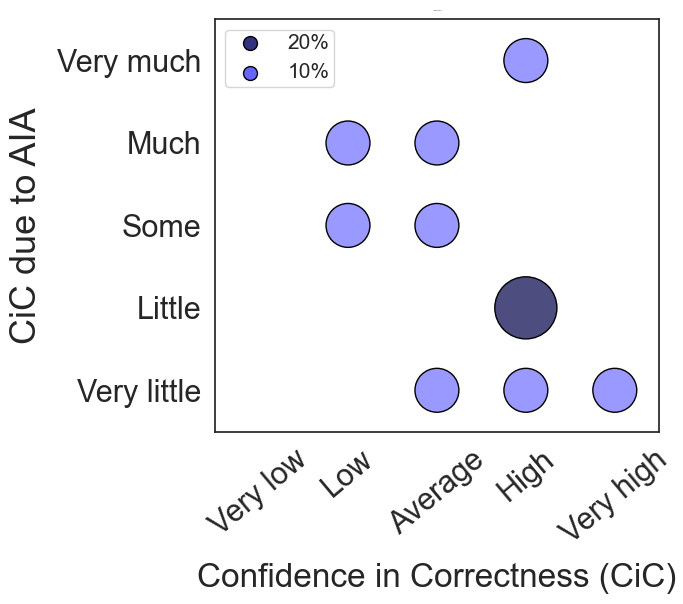} }}%
    \caption{Participants' confidence in correctness (CiC) of their B-specification with respect to how much of their level of confidence is due to the help of the AIA. Size of the bubble is proportional to the number of same answers.}
    \label{RQ2-qual}%
\end{figure}

Figure \ref{RQ2-qual} summarises the relationship of the answers given by participants in each iteration of the study. Participants who indicated that their performance was marginally (i.e. little or less) attributable to the help of the AIA are considered as participants who do not trust the AIA to write B-specifications. These participants are referred to as members of the \emph{distrustful group}. Conversely, participants who indicate that their performance was attributable to (some/much/very much) the help of the AIA are considered as participants who do trust the AIA to write B-specifications. These participants are referred to as members of the \emph{AI adopters} group. 

Using this group classification, iteration 1's results show that the distrustful group consisted of 67\% of the participants, whereas in iteration 2, it was 50\%. This shows that a large portion of the participants have acknowledged that their B-specifications were mainly their own production, with little contributions obtained from the AIA.

% Participants' ratings of their confidence in the correctness and helping of ChatGPT. Sizes represent the ratios of participants.

% BK
% \begin{figure*}[t!]
%     \centering
%     \begin{subfigure}[t]{0.5\textwidth}
%         \centering
%         %\includegraphics[height=1.2in]{a}
%         \includegraphics[height=6cm]{graphs/qual_partR1.png}
%         \caption{Iteration 1}
%     \end{subfigure}%
%     ~ 
%     \begin{subfigure}[t]{0.5\textwidth}
%         \centering
%         %\includegraphics[height=1.2in]{b}
%         \includegraphics[height=6cm]{graphs/qual_partR2.png}
%         \caption{Iteration 2}
%     \end{subfigure}
%     \caption{Participants' ratings of their confidence in the correctness and helping of ChatGPT. Sizes represent the ratios of participants.}
%     \label{R2-qual}
% \end{figure*}

\begin{figure}[t!]%
    \centering
    \subfloat[\centering Iteration 1]{{\includegraphics[width=3.9cm]{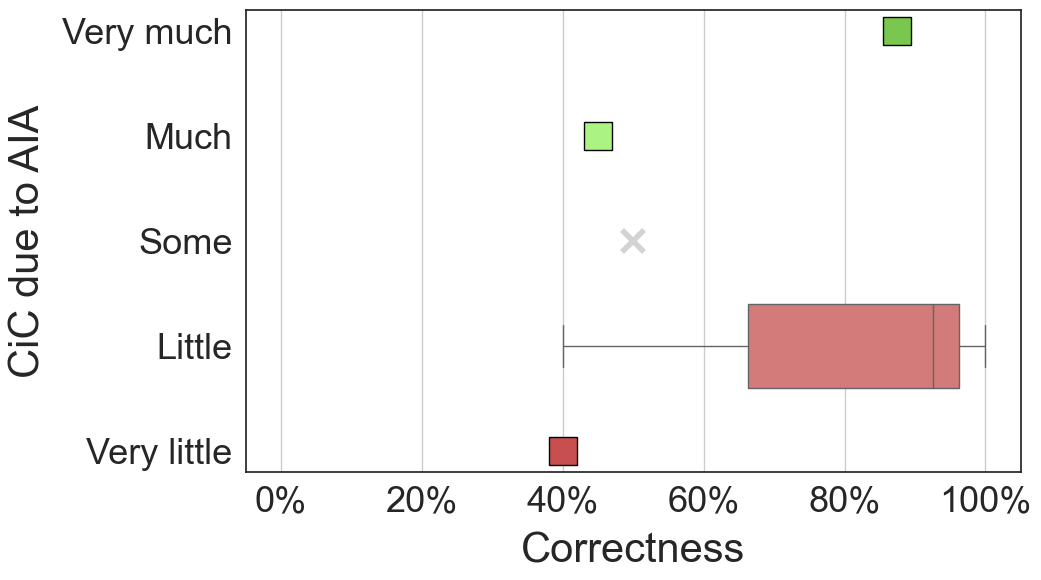} }}%
    \qquad
    \subfloat[\centering Iteration 2]{{\includegraphics[width=3.9cm]{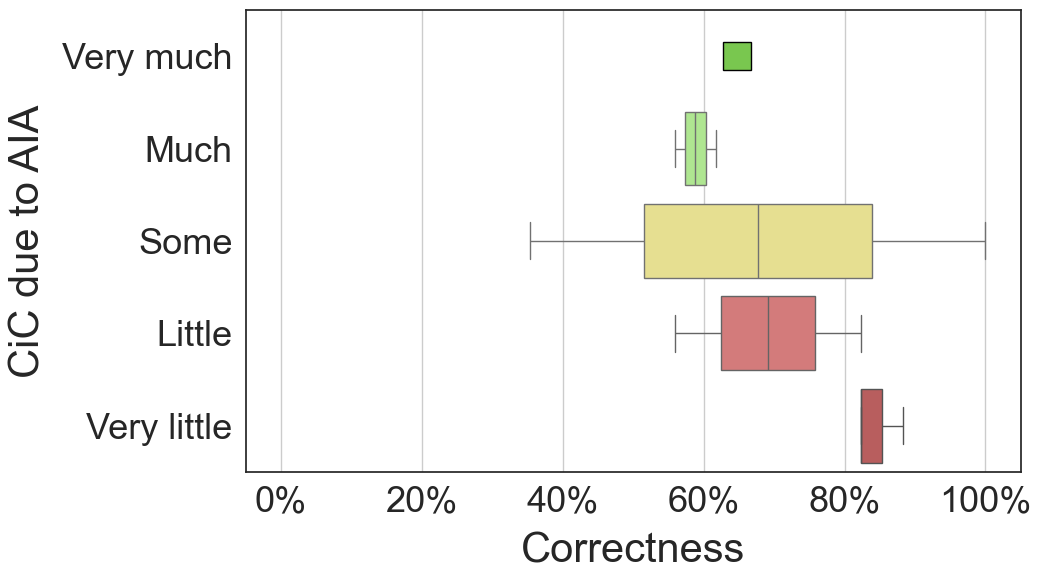} }}%
    \caption{Distribution of participant's level of confidence in correctness (CiC) is due to the help of the AIA placed over the actual correctness of their provided B-specification.}
    \label{RQ2-corr_trust}
\end{figure}

When comparing the participants' answers with respect to their actual obtained performance, (Figure \ref{RQ2-corr_trust}) we can see that those who indicated a marginal contribution from the AIA produced B-specifications with higher correctness levels. This trend is confirmed in the second iteration of the study. However, we noticed improvements in the level of correctness of participants who acknowledged having incorporated some AIA-generated answers into their B-specifications. 

Results show that \textbf{participants did not blindly trust the AIA to write B-specifications. Those who decided to use little of the AIA-generated answers have produced better B-specifications.}

% BK
% \begin{figure*}[t!]
%     \centering
%     \begin{subfigure}[t]{0.5\textwidth}
%         \centering
%         %\includegraphics[height=1.2in]{a}
%         \includegraphics[width=1\textwidth]{graphs/corr_trustR1.png}
%         \caption{Iteration 1}
%     \end{subfigure}%
%     ~ 
%     \begin{subfigure}[t]{0.5\textwidth}
%         \centering
%         %\includegraphics[height=1.2in]{b}
%         \includegraphics[width=1\textwidth]{graphs/corr_trustR2.png}
%         \caption{Iteration 2}
%     \end{subfigure}
%     \caption{Distribution of participant's perception on how much of their confidence in correctness (CiC) is due to the help of the AIA. These perceptions are reflected on the actual achieved correctness of the provided solutions.}
%     %\caption{The overall distribution of correctness of different participants' scores from Very Good to Very Poor among several properties of the specification: state variables, invariants, operations, other for each iteration. The results of the negative group with low variance are high, the results of the positive group are lower and have high variance for both iteration.}
%     \label{RQ2-corr_trust}
% \end{figure*}

\begin{figure}[t!]%
    \centering
    \subfloat[\centering Iteration 1]{{\includegraphics[width=3.9cm]{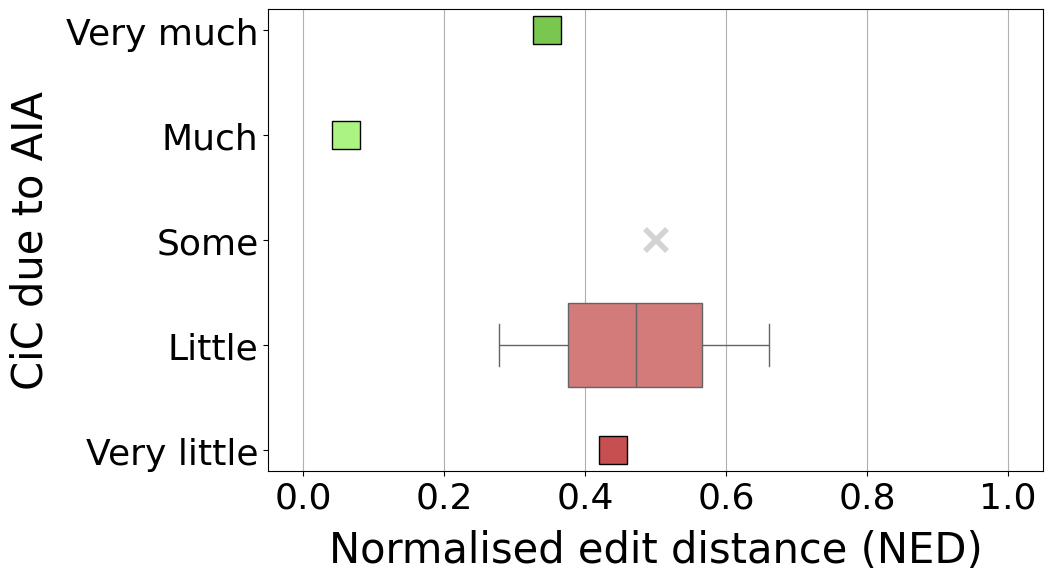} }}%
    \qquad
    \subfloat[\centering Iteration 2]{{\includegraphics[width=3.9cm]{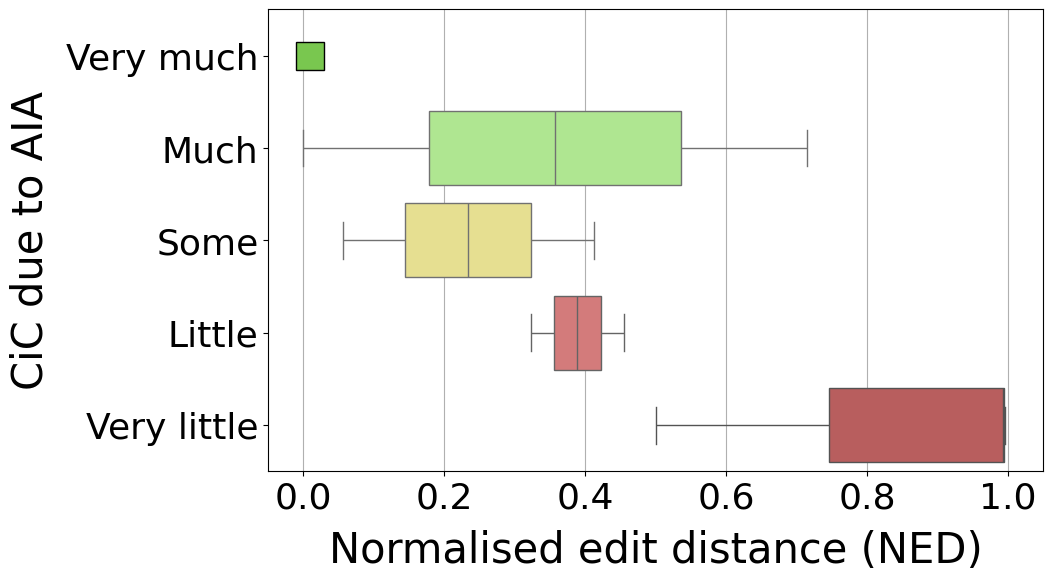} }}%
    \caption{Distribution of participant's level of confidence in correctness (CiC) is due to the help of the AIA placed over their normalised edit distance (NED).}
    \label{RQ2-eds_trust}
\end{figure}

In order to increase the reliability of these findings, we triangulated participants' answers to question \emph{q11} with the minimal NED between their B-specifications and AIA-generated answers from their dialogues. The results of this triangulation are summarised in Figure \ref{RQ2-eds_trust}. Recall that a high NED means low reuse of AIA-generated answers. In this study we use 0.4 as the threshold to determine if two strings are \textbf{similar} or not\footnote{If $ NED(s_{1},s_{2}) \in [0,0.4]$ then the two strings $s_{1}, s_{2}$ are considered similar.}

Results collected in both iterations show alignment between high NED and members of the \emph{distrustful group}. This allows us to conclude that the information provided by participants via the questionnaire match with their produced B-specifications. This matching increases the reliability of our conclusions.

\subsection{RQ3: How does students’ behaviour when interacting with the AIA affect the correctness of their B-specification?}\label{res-rq3}

% Analysis: 

% 1: prompts distribution
% how prompts are used among participants.
% This is descriptive information
% The variation in prompt \textit{strategies} among participants was studied and how such strategies affected correctness.

\begin{figure*}[t!]
    \centering
    \begin{subfigure}[t]{0.5\textwidth}
        \centering
        \includegraphics[width=1\textwidth]{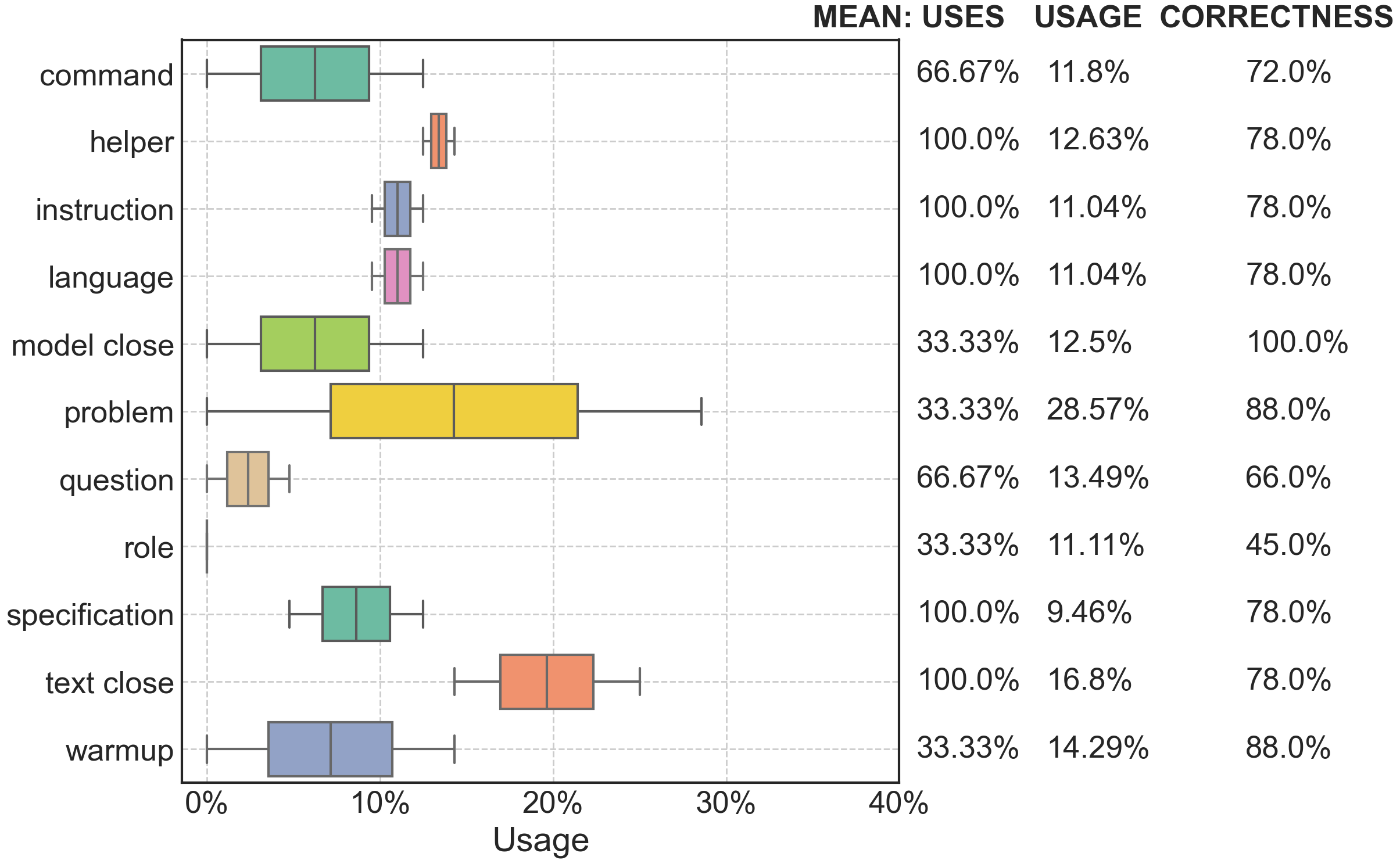}
        \caption{Iteration 1}
    \end{subfigure}%
    ~ 
    \begin{subfigure}[t]{0.5\textwidth}
        \centering
        \includegraphics[width=1\textwidth]{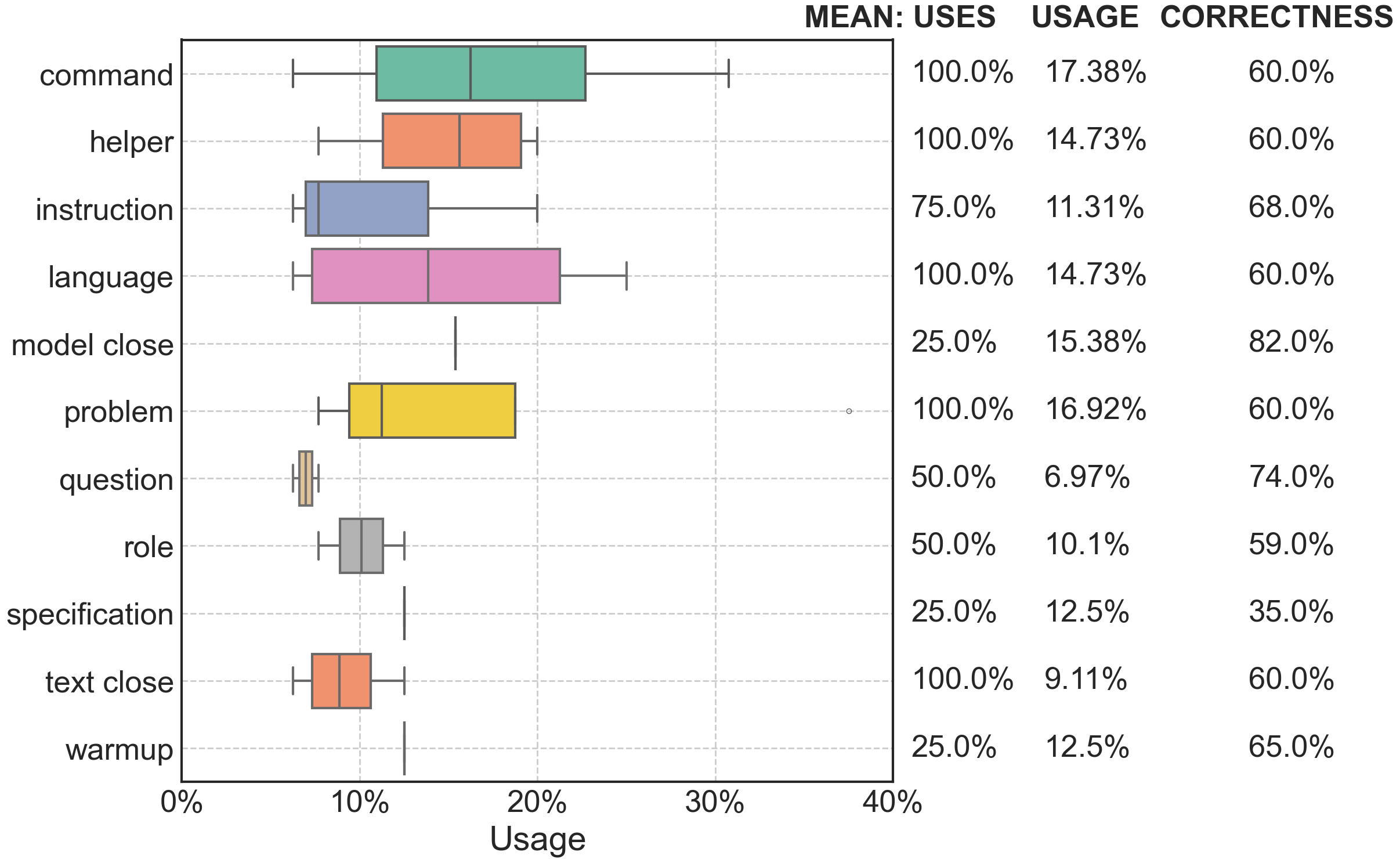}
        \caption{Iteration 2}
    \end{subfigure}
    \caption{How participants' prompt categories were distributed by \textit{usage} between the \textit{uses} of participants, and the average correctness level they got from involving these categories in dialogues.}
    \label{RQ3-prom_boxplot}

\end{figure*}

Participants used different strategies when interacting with the AIA. A strategy corresponds with the type of prompts (i.e. categories) used by participants, and how often prompts of a category were used during the duration of the interaction with the AIA (i.e. dialogue). Figure \ref{RQ3-prom_boxplot} shows the usage distribution of prompt categories among participants who provided ``good'' B-specifications and their NED was lower than 0.4. 

Iteration 1's prompt analysis reveals that \emph{helper, instruction, language, specification} and \emph{text close} are among the types of prompts used by all participants. This means, for instance, that every participant provided, at least once, to the AIA an example of a B-specification. Results show that on average 12.63\% of the prompts provided by participants were of type helper. Using helpers led to the production of B-specifications that on average had a correctness level of 78\%. 

Iteration 2's prompt analysis shows that its participants applied similar usage strategies. \emph{Helper, language} and \emph{text close} again appear among the types of prompts used by all participants. However, in this iteration \emph{command} and \emph{problem} prompts were also used, at least once, by all participants. Among them, the highest usage ratio corresponds with \emph{command} with 17.38\%. Participants who used prompts of this type produced B-specifications with an average correctness level of 60\%.

% 2: trends
% are prompt-trends similar between iterations

% 3: The impact of different prompting strategies on the correctness of AI-generated specification was analysed. 
% what are the prompts that help towards correctness
% is there a particular patter that may lead to increase the correctness of the specificaiton?

\begin{figure}[t!]%
    \centering
    \subfloat[\centering Iteration 1]{{\includegraphics[width=3.9cm]{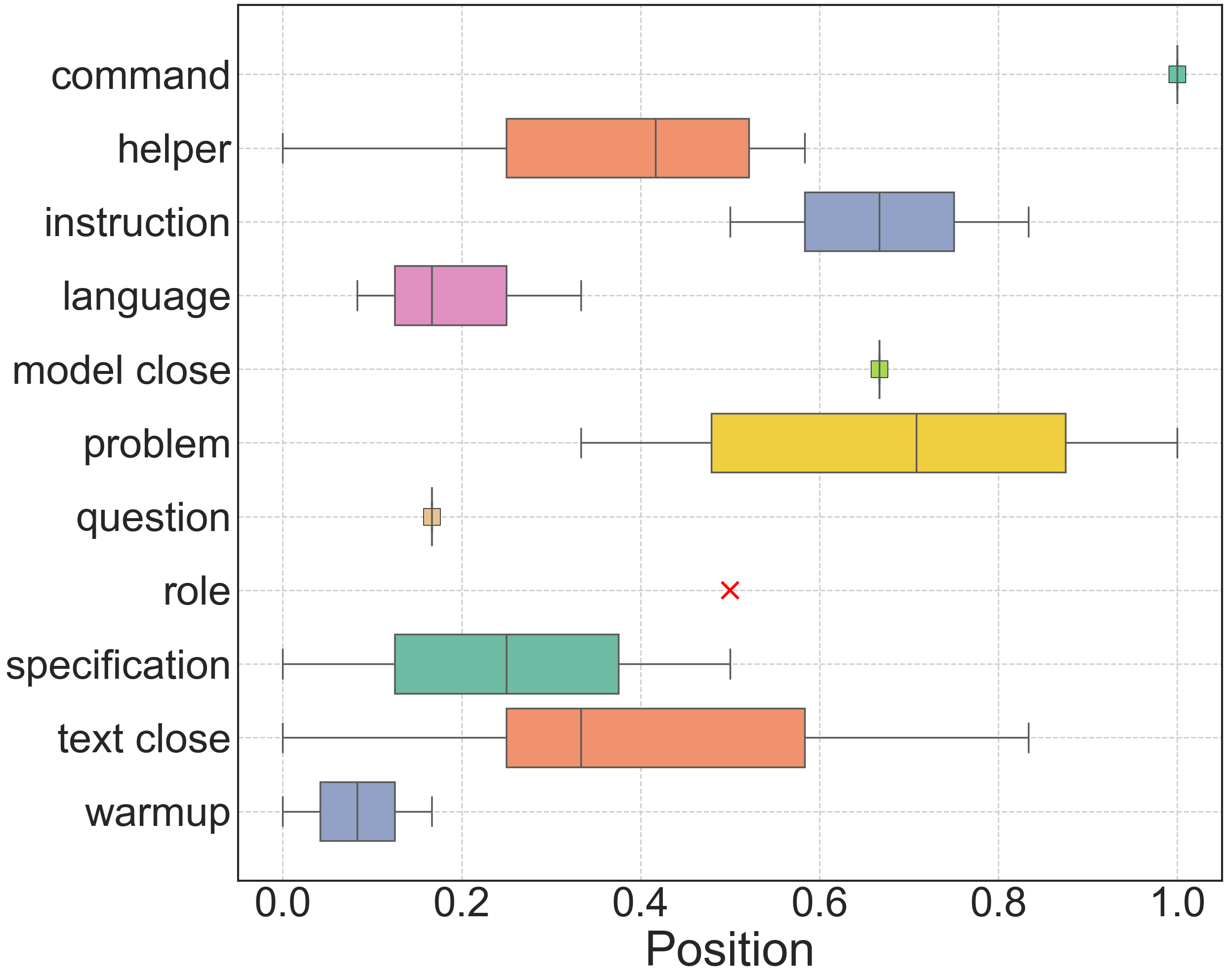} }}%
    \qquad
    \subfloat[\centering Iteration 2]{{\includegraphics[width=3.9cm]{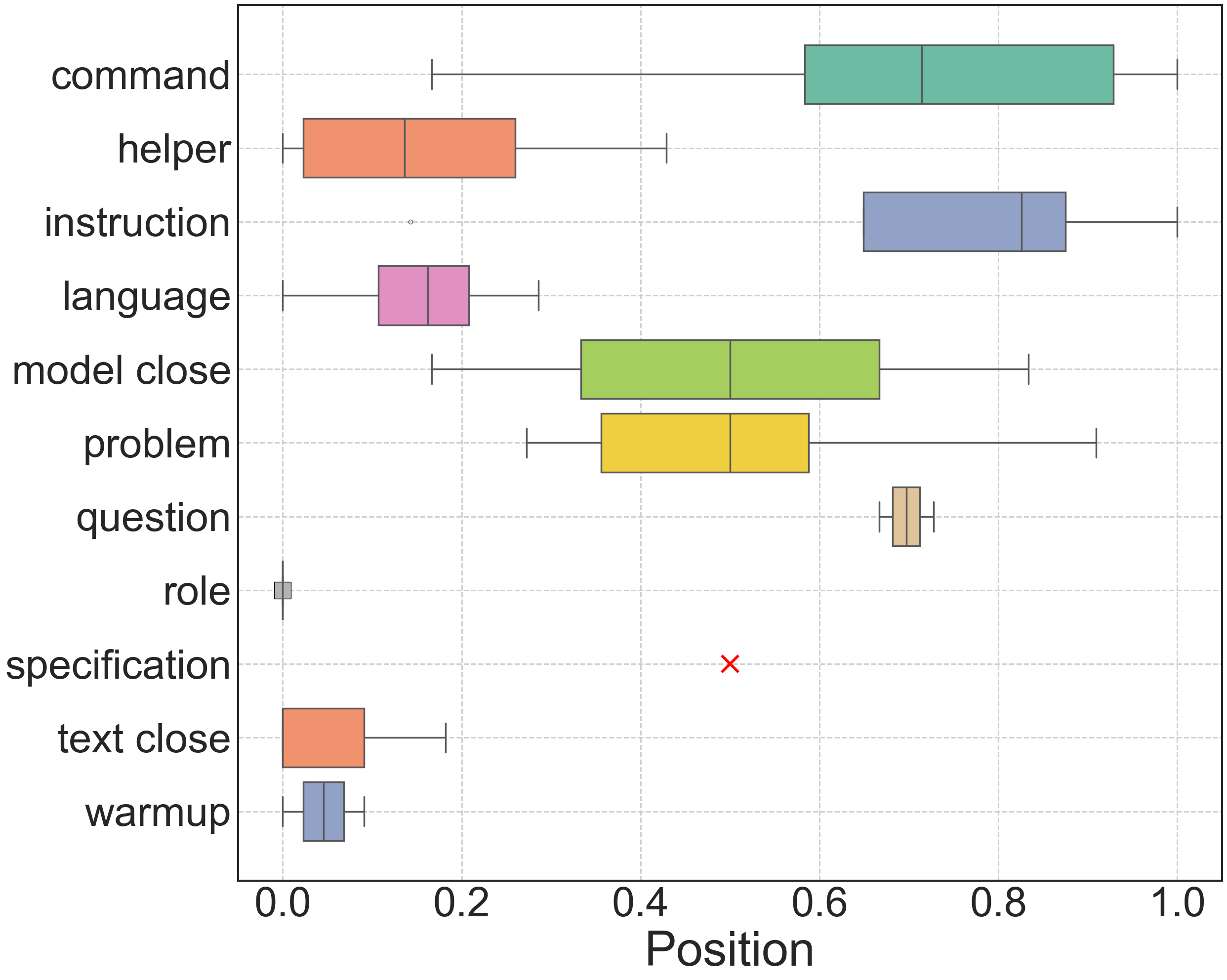} }}%
    \caption{Distribution of prompt categories with respect to their placement in the dialogue for participants who provided a ``good'' B-specification (i.e. correctness level $\ge$ 50\%).}
    \label{RQ3-boxplot-pos}
\end{figure}

While the usage distribution gives insights into how participants chose to interact with the AIA, it is also interesting to know when these types of prompts were used during the duration of the dialogue. Figure \ref{RQ3-boxplot-pos} shows how the different prompt categories are placed over the duration of the dialogue. Prompt categories placed in positions ranging between 0 and 0.3 are understood as prompts used at the beginning of the dialogue, whereas those with positions between 0.6 and 1 are more likely used towards the end. 

A behavioural pattern that appears in both iterations consists of \emph{warmup} prompts followed by  \emph{language}, and then \emph{instruction} interleaved with \emph{command} prompts towards the end. The usage of \emph{helper} prompts between \emph{language} and \emph{instruction} prompts is, to a certain extent, also observed. Based on these results, we can construct a possible \textbf{behavioural pattern with which to interact with the AIA to obtain ``good'' B-specifications. It may consist of (i) providing some context to the AIA, (ii) providing clear indications about the type of code to be generated, (iii) providing some helpers until an initial solution is obtained, then (iv) providing instructions and commands until a satisfactory AIA-generated answer is obtained.}

% Charts related to position

\begin{figure}[t!]%
    \centering
    \subfloat[\centering Iteration 1]{{\includegraphics[width=3.9cm]{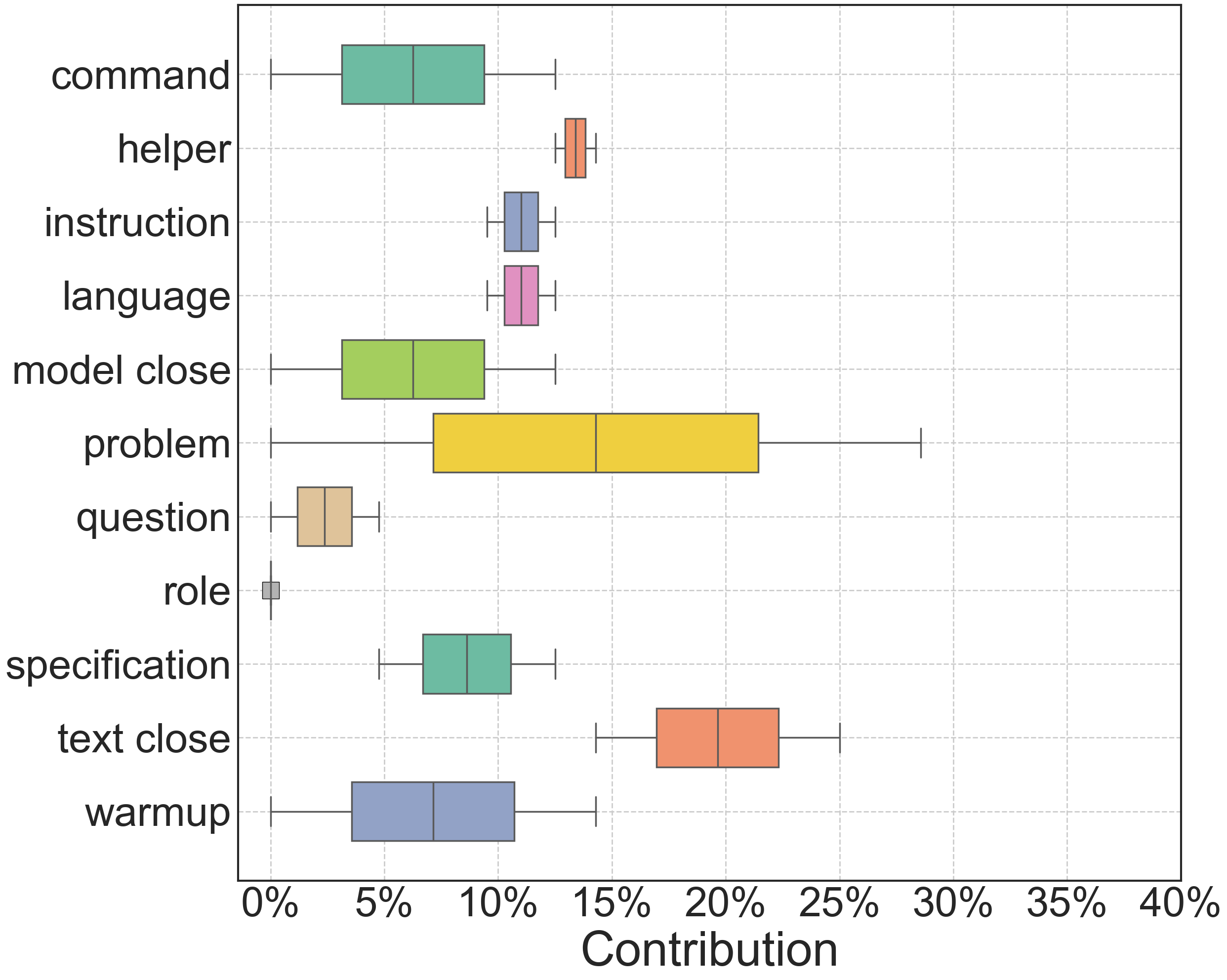} }}%
    \qquad
    \subfloat[\centering Iteration 2]{{\includegraphics[width=3.9cm]{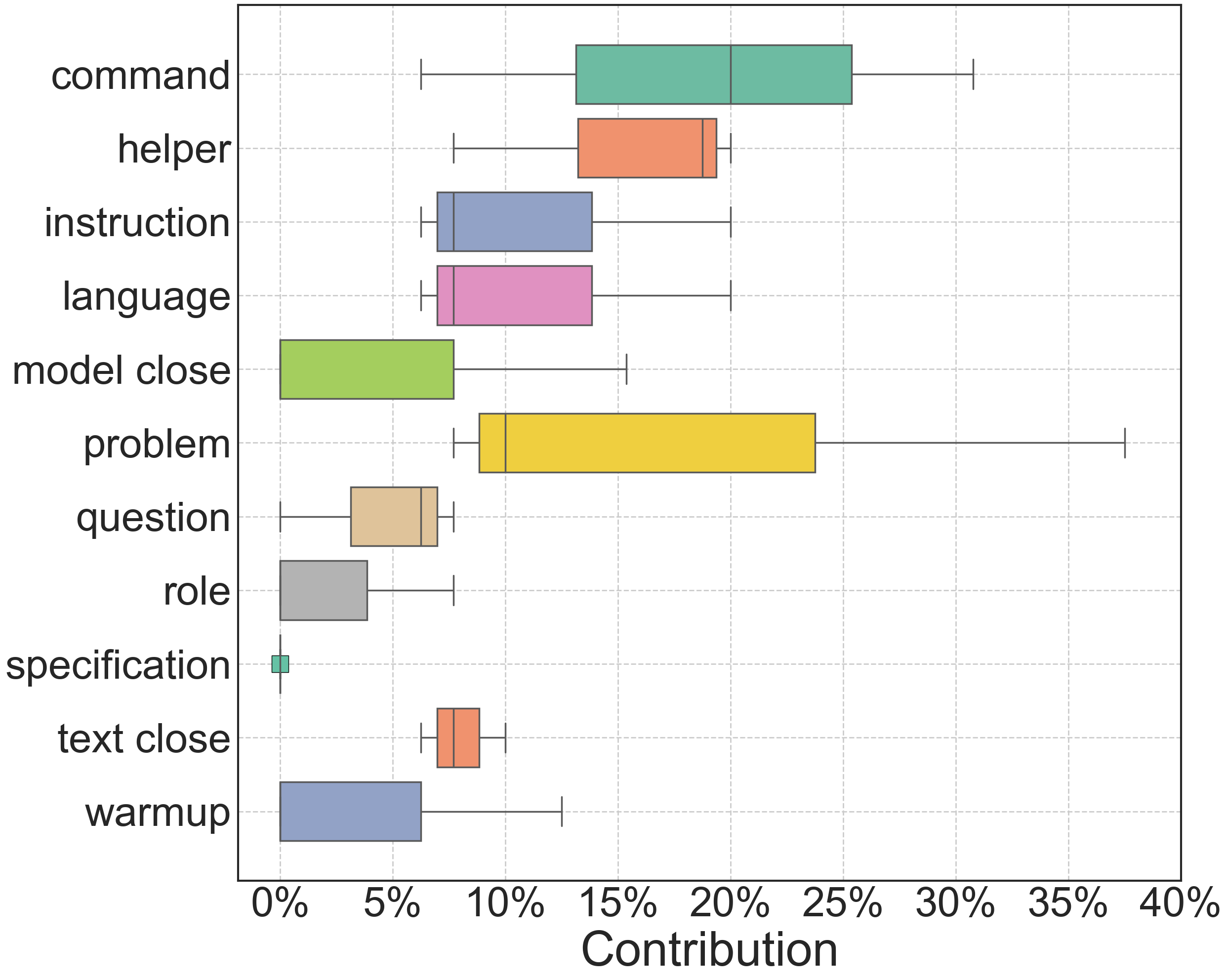} }}%
    \caption{Distribution of prompt categories with respect to their contribution to the participant's correctness for those who provided a ``good'' B-specification (i.e. correctness level $\ge$ 50\%).}
    \label{RQ3-boxplot-corr-pos}
\end{figure}

While identifying behavioural prompting patterns that help to produce ``good'' B-specifications is relevant, it is equally important to ascertain the impact that each prompt category may have on the correctness of the specification. We analysed the participants' interactions to determine how much each prompt category contributed to the correctness of the produced B-specifications.    

Comparing the results from iteration 2 with respect to those obtained from iteration 1, the categories whose contribution to the produced B-specification has increased are \emph{command, helper, question} and \emph{role}. Prompt categories whose contributions have considerably decreased are \emph{specification} and \emph{text close}. The decay of \emph{specification} prompts in the contributions towards correctness is an interesting finding. This confirms that the AIA does not generate B-specifications of the quality expected by the user. This evidence is consistent with the initial finding about the help offered by the AIA to write B-specifications. 

Inspecting the contributions of the categories that compose the behavioural pattern (Figure \ref{RQ3-boxplot-corr-pos}), we found a slight decay of \emph{warmups} (7.5\% vs. 6\% median values, iteration 1 and iteration 2, respectively), but an increase of \emph{helpers} (14\% vs. 18\%). \emph{Instructions} showed a large a decay (12\% vs. 7\%), whereas \emph{commands} largely increased (7\% vs. 20\%). \textbf{These findings support the overall contribution of the found behavioural pattern towards correctness. In particular, the interleaving of commands and instructions during the last part of the dialogue shows improvements in the achieved correctness level of B-specifications.}

\section{Discussion}

\subsection{Suggestions to educators}
The findings show that, as of today, the AIA cannot generate ``good'' B-specifications. The same findings, however, show that the AIA may provide some help during the process of creating the B-specification, but the correctness level of the final B-specification largely relies on the knowledge and skills of the student. This means that allowing students to use the AIA when writing B-specifications does not represent a risk for determining whether the expected learning outcomes have been acquired (assuming they are solely based on the correctness of the provided B-specification). Therefore, instructors are not required to invest time and effort into constructing complex assessment environments where access to the AIA is prohibited while other, necessary, third-party services remain accessible. Instructors may also use this information to warn students about the risks of blindly trusting AIA-generated answers without having mastered the concepts of the B-method and the B-language.

Findings also show that the level of help provided by the AIA is increasing. This means that it may eventually generate ``good'' B-specifications. Once this arrives, relying only on the correctness of the B-specification will no longer suffice for judging whether a student has acquired the expected learning outcomes. Instructors may want to run the study in their own course as a way of monitoring the progress of the AIA towards its assessment criterion. Knowing how close the AIA gets to a minimal passing mark may help them adjust the course's assessment criterion, the settings of the assessment environment, or both.

\subsection{Suggestions to students}

The findings provide insights into the types of prompts that contribute to obtaining ``good'' B-specifications. These insights also indicate at which moment certain types of prompts should be used to eventually reach a ``good'' B-specification. Needless to say, the benefits a student may gain from the AIA is directly proportional to the pre-acquired knowledge and developed skills on the B-method and B-language. 

The findings show that students who included portions of AIA-generated answers, either entirely or partially, in their B-specifications performed worse than those who used little to no AIA-generated text. This is evidence that the AIA should be used with care and mainly in settings in which a certain level of mastery has already been acquired.  

It is rare to obtain invariants that make the B-abstract machine consistent using the AIA. Additionally, findings show that students should not expect significant help with generating well-formed B-language expressions using the AIA. The same expectation applies when relying on the AIA to find the most appropriate B-operator (e.g. $\lhd$, the image antirestriction) required to write concise yet powerful predicates (e.g. $r := s \lhd r$).

% TODO
%\fred{Should we place a quotation given by a student via the post-activity survey?}

% \begin{tcolorbox}[colback=cyan!10!white, colframe=cyan!50!black, coltitle=black]
% \textbf{Participant R1.1005} gave the positive comment in the post-study survey: \texttt{``Chatgbt helps you getting a start with the assignment, and gives you ideas on how to define variables, sets, invariants etc``}
% \end{tcolorbox}

\section{Threats to validity}\label{sec:threats}

One important limitation of our study is the sample size of each iteration. Given the small sample size, each iteration of the study has low statistical power to detect smaller effects. This means that the AIA might actually have had a small effect on specifying correct B-specifications, but due to insufficient statistical power it was not detected. Due to the context where the study is embedded, it is quite unlikely to get a sample size large enough to detect small effects. Thus, rather than making decisions on evidence from a small single group, we decided to do two independent executions of the study to base our findings in trends appearing on both executions.

As sample sizes are not expected to grow, it would be hard to introduce a control group to the experiment. Without this group, it is difficult to conclude if the effects detected between the pretest and posttest are due to the use of the AIA. We made efforts to ensure that the same conditions apply (assessment criteria, complexity of the activities, and evaluators) to the pretest and posttest, such that effects may be attributed only to the treatment (i.e. use of the AIA). Nevertheless, it is worth mentioning that, so far, we have not found any effects from having allowed participants to use the AIA.

\section{Conclusion}

Our study highlights the limited impact of OpenAI's ChatGPT on undergraduate students' ability to write formal specifications using the B-method. While students expressed low trust in the AI, which correlated with better performance, some evidence suggests that ChatGPT may offer some support during the process that leads to obtaining the specification. However, the ultimate correctness of the B-specification largely depends on the students' knowledge and skills. This suggests that allowing students access to ChatGPT does not compromise their specification-based assessment. Future research should investigate the identified behavioural patterns in interactions with ChatGPT to evaluate its role in enhancing the correctness of B-specifications.

Moreover, as ChatGPT continues to improve, it may eventually produce correct B-specifications, making it necessary for educators to adapt their courses. In particular, their assessment strategies and environments.

\section{Data Availability}
Anonymised data\footnote{We have assigned an anonymous ID to each participant to ensure their anonymity.} and assets developed for their processing and analysis are made available in the replication package that accompanies this report~\cite{capozucca_2024_13914857}. This package should not only facilitate the replication of the results reported in this study, but also enable the execution of similar studies.

\bibliographystyle{IEEEtran}
\bibliography{main}

\end{document}